\documentclass[prb,twocolumn,showpacs]{revtex4-1}
\usepackage{amsmath}
\usepackage{graphicx}
\usepackage{bm}

\newcommand{\be}{\begin{equation}}
\newcommand{\ee}{\end{equation}}

\newcommand{\tb}{Tb$_2$Ti$_2$O$_7$}
\newcommand{\ho}{Ho$_2$Ti$_2$O$_7$}
\newcommand{\er}{Er$_2$Ti$_2$O$_7$}
\newcommand{\yb}{Yb$_2$Ti$_2$O$_7$}
\newcommand{\dy}{Dy$_2$Ti$_2$O$_7$}

\newcommand{\Journal}[4]{{\em #1} \textbf{#2}, #3 (#4)}
\newcommand{\PRev}{Phys.\ Rev. }

\begin{document}

\title{Exchange interactions in two-state systems: rare earth pyrochlores}

\author{S. H. Curnoe}
\email[Electronic address: ]{curnoe@mun.ca}
\affiliation{Department of Physics \& Physical Oceanography,
Memorial University of Newfoundland, St.\ John's, Newfoundland \& Labrador, 
A1B 3X7, Canada}

\begin{abstract}
The general form of the nearest neighbour exchange interaction for 
rare earth pyrochlores is derived based on symmetry.  
Generally, the rare earth angular momentum degeneracy is lifted
by the crystal electric field (CEF) into singlets and doublets.
When the CEF ground state is a doublet that is well-separated from the 
first excited state the CEF ground state doublet 
can be treated as a pseudo-spin of some kind.
The general form of nearest neighbour exchange interaction for pseudo-spins
on the pyrochlore lattice
is derived for three different types of pseudo-spins.  The methodology
presented in this paper can be applied to other two-state spin systems with a high
space group symmetry.
\end{abstract}

\maketitle

\section{Introduction}


Rare earth pyrochlores are crystals
that are famous for their unusual magnetic
correlations.  In all of these materials, the magnetic rare earth
(Gd, Tb, Dy, Ho, Er and Yb) 
ions are located  at the vertices of a network of corner-sharing 
tetrahedra, an example of an arrangement known as ``geometrical frustration."
In spite of their identical structures,
rare earth pyrochlores exhibit
a wide variety of states at
low temperatures,
including  ``spin ice" in
Ho$_2$Ti$_2$O$_7$ and Dy$_2$Ti$_2$O$_7$,\cite{harris1997,ramirez1999}  ``spin liquid" in
\tb,\cite{gardner2001}
and magnetic order in \er.\cite{champion2003}

Magnetic behaviour is generally modelled in terms of 
short range exchange interactions and 
longer range magnetic dipole-dipole interactions. 
The nearest-neighbour exchange interaction is anisotropic
in general and can be written as
\begin{equation}
H_{\rm ex} = \sum_{\langle ij \rangle} {\cal J}_{ij}^{\alpha\beta} J_i^{\alpha}J_j^{\beta},
\label{exchangeH}
\end{equation}
where $\langle ij \rangle$ are pairs of nearest neighbours at magnetic ion 
sites $i$ and $j$,
$\alpha, \beta = x,y,z$, and ${\cal J}_{ij}^{\alpha\beta}$ are phenomenological exchange
constants constrained by symmetry.
The dipole-dipole interaction takes the form
\begin{equation}
H_{\rm dipole} = \frac{1}{2} D a^3 \sum_{i,j} \frac{{\vec J}_i \cdot {\vec J}_j
}{|\vec{R}_{ij}|^3} - \frac{3(\vec{J}_i\cdot\vec{R}_{ij}) (\vec{J}_j \cdot \vec{R}_{ij})}{|\vec{R}_{ij}|^5},
\label{dipole}
\end{equation}
where  $D$ is the dipolar interaction strength, $a$ is the distance
between nearest neighbours, and $\vec{R}_{ij}$ is the displacement 
vector between sites $i$ and $j$. 
The nearest-neighbour 
part of the dipole-dipole 
interaction can be included in the exchange interaction by renormalising the exchange coupling constants. 
Although essential to the 
description of correlations in some systems (for example, in spin ices), longer range contributions to the dipole-dipole interaction are often neglected. 

In both (\ref{exchangeH}) and (\ref{dipole}) ${\vec J}_i$ is the total angular momentum (which shall be henceforth shortened to ``spin") of the magnetic ion at site $i$, and may
represent either the classical quantity or the quantum mechanical operator.
In the following, we will be concerned with quantum systems.

The rare earth ions have relatively large values of 
spin $J$  (as determined by Hund's rules) however the $2J+1$-fold degeneracy
is lifted by the 
crystal electric field (CEF) 
such that the energy levels are either  singlets or  doublets.  
In many of the pyrochlore magnets the splitting between the ground state
and first excited state large, of the order of 100 K, so that
magnetic properties can be modelled using only the CEF ground state,
neglecting the excited states.
The character table
of the group $D_{3}'$ (the relevant symmetry group of the CEF) is
shown in the appendix.
This table classifies spin states according to 
their transformation properties under the symmetry operations of $D_3'$.
According to the table, if $J$ is an integer then 
there are two different of singlets ($\Gamma_1$ and $\Gamma_2$)
 and a single ``non-Kramers" ($\Gamma_3$) doublet.
If $J$ is a half-integer then by Kramers' theorem all states are necessarily doublets, and there are two different
``Kramers" doublets, one of which transforms the same way as a $J=1/2$ spinor ($\Gamma_4$), and the other  not ($\Gamma_{5,6}$).  
Some examples of pyrochlores in the titanate family
with CEF ground state doublets are
\tb\ and \ho\ ($\Gamma_3$), \yb\ and \er\ ($\Gamma_4$)
and \dy\ ($\Gamma_{5,6}$).
In this article, we will review anisotropic nearest neighbour
exchange interactions for pyrochlores crystals where the magnetic ion ground states
are either Kramers or non-Kramers doublets.

The exchange interaction 
must be invariant under the symmetry operations
of the crystallographic space group, which, for the pyrochlore crystals, is 
$Fd\bar{3}m$. This space group has a face-centre cubic (fcc) Bravais lattice 
and its underlying point group is $O_h$ (octahedral).
The most general form of the  
exchange Hamiltonian between rare earth ions 
in the pyrochlore crystals has four independent terms,\cite{curnoe2008}
\begin{equation}
H_{\rm ex} = {\cal J}_1 X_1 + {\cal J}_2 X_2 + {\cal J}_3 X_3
+{\cal J}_4 X_4, 
\label{Hex}
\end{equation}
where ${\cal J}_i$ are four independent anisotropic exchange 
constants.

It is convenient to write the exchange terms $X_i$ using a set of local axes,
defined  such that the
local $z$-axis (the spin quantisation axis) is the 3-fold symmetry axis of
the CEF at the rare earth site (see Fig.\ \ref{fig1} in the Appendix).
This coordinate system is defined in Appendix A.
Using this notation, the exchange terms are
\begin{eqnarray}
X_1 & = & -\frac{1}{3} \sum_{\langle ij\rangle} J_{iz}J_{jz} \label{x1} \\
X_2 & = & -\frac{\sqrt{2}}{3} \sum_{\langle ij\rangle} [\Lambda_{ij}(J_{iz}
J_{j+} + J_{jz}J_{i+}) + {\rm h.c.}] \label{x2} \\
X_3  & = & \frac{1}{3} \sum_{\langle ij \rangle } (\Lambda_{ij}^{*}
J_{i+}  J_{j+} + {\rm h.c.} ) \label{x3}  \\
X_4 & = & -\frac{1}{6} \sum_{\langle ij \rangle }(J_{i+} J_{j-}  + {\rm h.c.}),
\label{x4}
\end{eqnarray}
where h.c. stands for ``Hermitian conjugate" and $J_{\pm}= J_x\pm i J_y$. 
The rare earth ions are found at the 16d Wyckoff position of the space group
$Fd\bar{3}m$, which has four inequivalent sites, therefore each rare earth ion site $i$ or $j$
can be specified by a fcc lattice vector and a site number, 1, 2, 3 or 4.
Sites which are nearest neighbours will either belong to the same lattice vector or will 
differ by a single fcc translation.
The complex coefficients $\Lambda_{ij}$ depend only on the site numbers, and not the lattice vectors.  They are  
$\Lambda_{12} = \Lambda_{34} =1$ and  $\Lambda_{13} = \Lambda_{24}
= \Lambda_{14}^{*} = \Lambda_{23}^{*} = \varepsilon \equiv 
\exp \left(\frac{2\pi i}{3}\right)$.
The numerical coefficients in front of each term in (\ref{x1}-\ref{x4}) are selected so that the sum
over all four terms is just the isotropic exchange interaction
\begin{equation}
X_1+X_2+X_3+X_4 = \sum_{\langle ij\rangle} \vec{J}_i \cdot \vec{J}_j.
\end{equation} 
Several different forms of $H_{\rm ex}$ have appeared in the literature which amount to 
different linear combinations of the exchange terms $X_i$. 
Some alternative forms are given in Appendix B.

To prove that $H_{\rm ex}$ (Eq.\ \ref{Hex}) is the most general form of the exchange interaction we must demonstrate the following: {\em i)} that there are exactly 
four independent terms that are bilinear in the spin
operators, {\em ii)} that the four terms (\ref{x1}-\ref{x4}) are in fact independent and {\em iii)} invariant 
under the space group operations and time reversal.
In order to prove any of these assertions, one needs to know how each 
spin operator transforms under the operations of $O_h$, the point group 
associated with $Fd\bar{3}m$.  The transformations of the spin operators
are listed in Appendix A, 
along with the calculation that shows that there are four independent terms in the 
nearest neighbour exchange interaction.  It is
obvious that (\ref{x1}-\ref{x4}) are independent because they contain different 
kinds of operators,
and their invariance can be checked using the transformations given in 
Appendix A.

This paper is concerned with nearest neighbour interactions between
each of the three different types of CEF ground state
doublets ($\Gamma_3$, $\Gamma_4$, and $\Gamma_{5,6}$).
We begin by finding the restriction of the exchange interaction $H_{\rm ex}$ 
(\ref{Hex}) to the CEF ground state
doublet.
Clearly the restriction of $H_{\rm ex}$ will
be non-trivial only when there are non-vanishing matrix elements for
$J_z$ and for 
$J_{\pm}$ within the restriction.  Generally, $J_z$ has non-vanishing matrix elements for each of the three kinds of doublet, however
$J_{\pm}$ is only non-vanishing for a $\Gamma_4$ doublet:  symmetry requirements
(including time reversal) force the matrix elements for $J_{\pm}$ to vanish within $\Gamma_3$ and $\Gamma_{5,6}$ doublets.
Therefore when $H_{\rm ex}$ is restricted 
to a $\Gamma_3$ or a $\Gamma_{5,6}$ subspace the
model contains no quantum mechanical effects because the matrix elements of the
$J_\pm$ operators vanish within either of these restrictions.  
However, because a spin-1/2 spinor belongs to the $\Gamma_4$ representation, when $H_{\rm ex}$ is restricted to a $\Gamma_4$ doublet the resulting model 
will be the $J=1/2$ version of $H_{\rm ex}$ (\ref{Hex}).

Physically, there are other interactions which may be relevant
besides the exchange interaction that can generate 
non-zero matrix elements between the states in 
$\Gamma_3$ or $\Gamma_{5,6}$ doublets. 
Two that have been previously studied are {\em i)}
mixing with higher CEF levels,\cite{molavian2007,curnoe2013} and {\em ii)} higher multipole (quadrupolar or octupolar) interactions.\cite{princep2013,xu2015,petit2016,benton2016} 
These two types of interactions are handled differently.

The relevance of 
excited state CEF levels in \tb\ was recognised long ago, in the first studies 
of magnetostriction.\cite{aleksandrov1985,mamsurova} In \tb, the ground state and first excited CEF levels are both
$\Gamma_3$ doublets, separated by an energy difference of only
about 18 K.  The first excited state admixes to the ground state to second order in
perturbation theory 
via the exchange interaction, and gives rise to non-vanishing $J_{\pm}$ matrix elements
which are recorded in the intensity patterns of 
diffuse neutron scattering
experiments.\cite{molavian2007,curnoe2013}  
To model these interactions we make use of a map
between $\Gamma_3$ doublets and spin-1/2 ($\Gamma_4$) 
doublets. If the rare earth spins
are considered in groups of four (the four sites
 of a tetrahedron) there is an exact symmetry match
between $\Gamma_3$ doublets and $\Gamma_4$ doublets.\cite{curnoe2013}  This map allows one to write to write the 
interactions between the states in a 
$\Gamma_3$  doublet as an effective spin-1/2 exchange interaction;
{\em i.e.} the effective Hamiltonian is the $J=1/2$ version of
(\ref{Hex}),  with all
four coupling constants ${\cal J}_i$ of the effective $J=1/2$ model   
non-zero.
There is no similar map for 
$\Gamma_{5,6}$ doublets.  


On the other hand, higher multipole interactions have been invoked to model 
praseodymium pyrochlores with $\Gamma_3$ CEF ground states, such as Pr$_2$Zr$_2$O$_7$\cite{petit2016} and 
Pr$_2$Sn$_2$O$_7$,\cite{princep2013}
since quadrupolar interactions are the lowest order interactions which
yield non-zero $J_{\pm}$ matrix elements within 
the $\Gamma_3$ restriction (no mixing with excited CEF levels is required).
Similarly, octupole interactions can yield non-zero matrix elements for
$\Gamma_{5,6}$ states, as in Nd$_2$Zr$_2$O$_7$.\cite{huang2014,xu2015,benton2016,petit2016b}
This article describes how to cast these higher multipole interactions as
effective exchange interactions.

In the following sections we use symmetry methods to derive 
general effective nearest neighbour exchange interaction 
models for each of the three different kinds of CEF 
ground state doublets, $\Gamma_4$,  $\Gamma_{5,6}$ 
and $\Gamma_3$.

\section{Exchange interaction for $\Gamma_4$ CEF
ground states}

We first consider 
$H_{\rm ex}$
(Eq.\ \ref{Hex}) for  $J=1/2$, using an alternate approach, 
as follows.  
The components of the spin operator $\vec S$ are {\em defined}
in terms of the spin-1/2 states $|\pm\rangle$,
\begin{eqnarray}
S_{x}  & = & |+\rangle \langle -|+ |-\rangle\langle +| \label{Jjx}\\
S_{y} & = &  -i|+\rangle \langle -| +i |-\rangle\langle +| \label{Jjy}\\
S_{z} & = & |+\rangle \langle +| - |- \rangle \langle -| \label{Jjz}
\end{eqnarray}
The quantisation axis (the $z$-axis) implied by this notation
points in the direction of the 3-fold symmetry axis of the rare earth site
(the local $z$-axis).
The transformation properties of ${\vec S}$ under rotations and time reversal follow from
the transformation
properties of the spin-1/2 bras and kets, 
where the rotation operator is given by $\exp(-i \theta \vec J \cdot \hat{n})$
for $J=1/2$.
We find that ${\vec S}$ transforms the same way as 
the angular momentum ${\vec J}$ (as it must), 
and so
the general form of 
the exchange interaction for ${\vec S}$ is the $J=1/2$ version of
$H_{\rm ex}$ (\ref{Hex}).
We also note that under $D_3'$, $S_{z}$ transforms as $\Gamma_2$, while 
$S_{x,y}$ transforms as $\Gamma_3$, and $\vec S$ changes sign under time
reversal.

More generally, any $\Gamma_4$ doublet, such as the CEF ground state of 
erbium in \er,
\begin{widetext}
\begin{equation}
|\Gamma^{\pm}_{4({\rm Er})}\rangle = 0.471|\pm 13/2\rangle \pm 0.421|\pm 7/2\rangle - 0.569 |\pm 1/2\rangle \mp 0.240 |\mp 5/2\rangle 
+ 0.469 |\mp 11/2\rangle 
\end{equation}
\end{widetext}
can be used to define operators analogous to (\ref{Jjx}-\ref{Jjz}),
\begin{eqnarray}
\tilde{S}_x & = &  |\Gamma^{+}_4\rangle \langle \Gamma^{-}_4|+ |\Gamma^{-}_4\rangle\langle \Gamma^{+}_4| \label{S4x}\\
\tilde S_{y} & = &  -i|\Gamma^{+}_4\rangle \langle \Gamma^{-}_4| +i |\Gamma^{-}_4\rangle\langle \Gamma_4^{+}| \\
\tilde S_{z} & = & |\Gamma^{+}_4\rangle \langle \Gamma^{+}_4| - |\Gamma^{-}_4 \rangle \langle \Gamma^{-}_4| \label{S4z}
\end{eqnarray}
These operators transform in the same way as $\vec S$
under the $D'_3$ group operations. Moreover, these operators can be 
scaled by appropriate factors so that
they have the same eigenvalues as $\vec S$ (i.e., $\pm \hbar/2$):  
$\tilde S_{x,y} =  t S_{x,y}$ and $\tilde S_z = j S_z$, where
$t = \frac{\langle \Gamma^{+}_4 |J_{+}|\Gamma_4^{-}\rangle}{\hbar}$ and
$ j = \frac{\langle \Gamma^{+}_4 |J_{z}|\Gamma_4^{+}\rangle}{\hbar/2}$.
Therefore the general form of the exchange interaction using such operators will 
be the $J=1/2$ version of $H_{\rm ex}$ (\ref{Hex}) with the constants $j$ and $t$ absorbed into
the exchange constants ${\cal J}_i$.
This Hamiltonian has been used to model Yb$_2$Ti$_2$O$_7$,\cite{ross2011,thompson2011,yan2017}
Er$_2$Ti$_2$Ti$_7$,\cite{mcclarty2009,chapuis2010,savary2012b,bonville2013,zhitomirsky2014,petit2014,yan2017}
and Er$_2$Sn$_2$O$_7$.\cite{guitteny2013,yan2017}

\section{Exchange interaction for $\Gamma_{5,6}$ CEF ground states}
Examples of rare earth pyrochlores with a $\Gamma_{5,6}$ CEF ground state
include Dy$_2$Ti$_2$O$_7$,\cite{bertin2012} Nd$_2$Ir$_2$O$_7$,\cite{watahiki2011} and Nd$_2$Zr$_2$O$_7$.\cite{xu2015}
The CEF ground 
state of dysprosium in \dy\ is\cite{bertin2012}
\begin{widetext}
\begin{equation}
|\Gamma_{5,6 ({\rm Dy})}^{\pm}\rangle = 0.981|\pm 15/2\rangle \pm 0.190 |\pm 9/2\rangle - 0.022|\pm 3/2\rangle
\mp 0.037|\mp 3/2\rangle + 0.005 |\mp 9/2\rangle \pm 0.001|\mp 15/2\rangle.
\end{equation}
\end{widetext}
In this rendering of the doublet the coefficients are 
real 
 and the matrix elements of
$J_z$ within this doublet are 
$$\left(\begin{array}{cc}
7.379 & 0.005 \\
0.005 & -7.379\end{array} \right).
$$   
That is, within this restriction, $J_z$ can be represented by the matrix
$7.379 \sigma_z + 0.005 
\sigma_x$.  
The matrix elements for $J_{x}$ and $J_y$ are zero within this restriction,
and
we find that quadrupolar operators of the 
form $J^{\alpha}J^{\alpha}$ are proportional to the identity, while
$J^{\alpha}J^{\beta}$ ($\alpha \neq \beta$) vanish.
The non-zero octupolar moments are $J_x^3$, $J_z^3$, $J_x J_y^2$ and $J_y^2 J_z$,
proportional to combinations of $\sigma_x$ and $\sigma_z$, and 
$J_y^3$ and $J_x^2 J_y$, proportional to $\sigma_y$.

We can define a set of operators based on these kets similar to (\ref{Jjx}-\ref{Jjz})
 and (\ref{S4x}-\ref{S4z}), 
\begin{eqnarray}
\tau_{x}  & = & |\Gamma_{5,6}^{+}\rangle \langle \Gamma_{5,6}^{-}|+|\Gamma_{5,6}^{-}\rangle\langle \Gamma_{5,6}^{+}|  \\
\tau_{y} & = &  -i|\Gamma_{5,6}^{+}\rangle \langle \Gamma_{5,6}^{-}| +i |\Gamma_{5,6}^{-}\rangle\langle \Gamma_{5,6}^{+}|   \\
\tau_{z} & = & |\Gamma_{5,6}^{+}\rangle \langle \Gamma_{5,6}^{+}| - |\Gamma_{5,6}^{-} \rangle \langle \Gamma_{5,6}^{-}|. 
\end{eqnarray}
The $\tau_{\alpha}$ are represented by the matrices $\sigma_i$ in the $\{|\Gamma_{5,6}^{+}\rangle,
|\Gamma_{5,6}^{-}\rangle\}$ basis.  
Therefore, the actions of $J_z$, $J_x^3$, $J_z^3$, $J_x J_y^2$ and $J_y^2 J_z$ 
in the restricted space of the $\Gamma_{5,6}$ doublets are
equivalent to linear combinations of $\tau_x$ and $\tau_z$, while $J_y^3$ and $J_x^2 J_y$
are proportional to $\tau_y$. In other words, the operator
$\tau_y$ is only present in models which include octupole moments.

The doublet can also be 
written as 
\begin{eqnarray}
|\Gamma_5\rangle & = &  (|\Gamma_{5,6}^{+}\rangle + i |\Gamma_{5,6}^{-}\rangle)/\sqrt 2 \\
|\Gamma_6\rangle & = &  (|\Gamma_{5,6}^{+}\rangle - i |\Gamma_{5,6}^{-}\rangle)/\sqrt 2.
\end{eqnarray}
The rendering of the doublet as $\{|\Gamma_{5,6}^{+}\rangle, |\Gamma^{-}_{5,6}\rangle\}$ has the physical interpretation
of a spin that points into ($|\Gamma_{5,6}^{+}\rangle$) or out of ($|\Gamma_{5,6}^{-}\rangle$) a tetrahedron, while 
$\{|\Gamma_5\rangle, |\Gamma_6\rangle\}$ is a pair of states that are time-reversed partners 
which transform according to two separate representations,
$\Gamma_5$ and $\Gamma_6$, in Table \ref{D3}.
Either rendering is a valid basis for the space spanned by these kets.

Let us define pseudo-spin operators based on the kets $\{|\Gamma_5\rangle,|\Gamma_6\rangle\}$, 
\begin{eqnarray}
\beta_{x}  & = & |\Gamma_5\rangle \langle \Gamma_6|+ |\Gamma_6\rangle\langle \Gamma_5| = \tau_z\\
\beta_{y} & = &  -i|\Gamma_5\rangle \langle \Gamma_6| +i |\Gamma_6\rangle\langle \Gamma_5|= \tau_x \\
\beta_{z} & = & |\Gamma_5\rangle \langle \Gamma_5| - |\Gamma_6 \rangle \langle \Gamma_6| = \tau_y.
\end{eqnarray}
The transformation of these operators under the  operations of the 
point group $D'_3$ is  determined
by the transformation of their constituent bras and kets under rotations.
It is found that 
$\beta_x$ and $\beta_y$ transform as $\Gamma_2$, while $\beta_z$ transforms as $\Gamma_1$, and
all components of ${\vec \beta}$ change sign under time reversal.
The transformations of these operators under the space group operations are given 
in Table \ref{spintable}) of Appendix A.
By analysing the decomposition of the representation generated by bilinears in $\vec \beta$,
we find that 
there are exactly four exchange terms allowed by the space group symmetry of 
the pyrochlore lattice, with slightly different forms compared
to the $\Gamma_4$ case:\cite{huang2014}
\begin{eqnarray}
X_{\Gamma_{5,6},1}  & = &  
\sum_{\langle ij\rangle}  \beta_{iz}\beta_{jz} \label{betaz}\\
X_{\Gamma_{5,6},2}  & = & \sum_{\langle ij\rangle}  \beta_{ix}\beta_{jx}  \\
X_{\Gamma_{5,6},3}  & = & \sum_{\langle ij\rangle}  \beta_{iy}\beta_{jy}  \\
X_{\Gamma_{5,6},4}  & = & \sum_{\langle ij\rangle}  (\beta_{ix}\beta_{jy} + \beta_{iy}\beta_{jx})/2 \label{betaxy}
\end{eqnarray}
The total Hamiltonian therefore has four terms,
\begin{equation} 
H  =  J_1 X_{\Gamma_{5,6},1} + J_2 X_{\Gamma_{5,6},2} +J_3 X_{\Gamma_{5,6},3} +J_4 X_{\Gamma_{5,6},4}.   \label{HGamma5,1}
\end{equation}
However,  last three terms in $H$ can be replaced by two terms,\cite{huang2014}
\begin{eqnarray}
\tilde{X}_{\Gamma_{5,6},2}  & = & \sum_{\langle ij\rangle}  \tilde{\beta}_{ix}\tilde{\beta}_{jx}  \\
\tilde{X}_{\Gamma_{5,6},3}  & = & \sum_{\langle ij\rangle}  \tilde{\beta}_{iy}\tilde{\beta}_{jy} , 
\end{eqnarray}
where $\tilde{\beta}_{x} = \cos \theta \beta_{x} + \sin \theta \beta_{y}$
and $\tilde{\beta}_y = -\sin \theta \beta_{x} + \cos \theta \beta_{y}$
are the operators rotated in the local $xy$ plane. 
Then the Hamiltonian (\ref{HGamma5,1}) is equivalent to
\begin{equation}
 H =  J_1 X_{\Gamma_{5,6},1} + \tilde{J}_2 \tilde X_{\Gamma_{5,6},2} +\tilde{J}_3 \tilde X_{\Gamma_{5,6},3}, 
\label{HGamma5,2}
\end{equation}
where
\begin{eqnarray}
\tilde{J}_2 &  = & \frac{J_2+J_3 + \sqrt{(J_2-J_3)^2 +  J_4^2}}{2}\\
\tilde{J}_3 & = &  \frac{J_2+J_3 - \sqrt{(J_2-J_3)^2 +  J_4^2}}{2}\\ 
\tan 2 \theta & = & \frac{ J_4}{J_2 - J_3}.
\end{eqnarray}
This three-parameter model has been  applied to Nd$_2$Zr$_2$O$_7$.\cite{benton2016,petit2016b}
As noted in Ref.\ \onlinecite{benton2016}, the phase diagram of the exchange model (\ref{HGamma5,1}) or  (\ref{HGamma5,2}) does not depend on the 
angle $\theta$, however the response of the system to an 
applied magnetic field is sensitive to this angle 
because it corresponds to a rotation in the local $xz$ plane of each 
rare earth magnetic 
moment.

Considering the model on a single tetrahedron yields further insight about
the symmetry of the spin states.
On a single tetrahedron the $\Gamma_{5,6}$ doublet space is spanned by sixteen kets of the
form $|\alpha_1 \alpha_2 \alpha_3 \alpha_4 \rangle
\equiv |\Gamma_{5,6}^{\alpha_1}\rangle_1 \otimes 
|\Gamma_{5,6}^{\alpha_2}\rangle_2 \otimes |\Gamma_{5,6}^{\alpha_3}\rangle_3 
\otimes|\Gamma_{5,6}^{\alpha_4}\rangle_4 $, where $\alpha_i = \pm$ represents the
CEF state of the site $i$ ($i = 1,2,3,4$) of a tetrahedron.  Linear combinations of these kets
are basis functions belonging to the 
representations $3 A_1\oplus 2 A_2 \oplus E \oplus 
2 T_1 \oplus T_2$ of the point group $T_d$, the symmetry group of a tetrahedron: 
\begin{widetext}
\begin{eqnarray}
|A_1^{(1)}\rangle & = &  (|++++\rangle + |----\rangle)/\sqrt{2} \\
|A_1^{(2)}\rangle & = & (|++--\rangle +|--++\rangle + |+-+-\rangle
+|-+-+\rangle + |-++-\rangle + |+--+\rangle)/\sqrt{6} \\
|A_1^{(3)}\rangle & = & (|+++-\rangle +|++-+\rangle +|+-++\rangle
-|---+\rangle - |--+-\rangle - |-+--\rangle - |+---\rangle )/\sqrt{8} \\
|A_{2}^{(1)}\rangle & = &   (|++++\rangle - |----\rangle)/\sqrt{2} \\
|A_2^{(2)}\rangle & = &  (|+++-\rangle +|++-+\rangle +|+-++\rangle
+|---+\rangle + |--+-\rangle + |-+--\rangle + |+---\rangle )/\sqrt{8} \\
|E_1\rangle & = & (|++--\rangle + |--++\rangle + |+-+-\rangle 
+|-+-+\rangle - 2(|-++-\rangle + |+--+\rangle))/\sqrt{12} \\
|E_2\rangle & = & (|++--\rangle + |--++\rangle -( |+-+-\rangle + |-+-+\rangle))/\sqrt{4} \\
|T_{1x}^{(1)}\rangle & = & (|+--+\rangle - |-++-\rangle)/\sqrt{2}\\
|T_{1y}^{(1)}\rangle & = &  (|+-+-\rangle - |-+-+\rangle)/\sqrt{2} \\
|T_{1z}^{(1)}\rangle & = & (|++--\rangle - |--++\rangle)/\sqrt{2} \\
|T_{1x}^{(2)}\rangle & = & (-|+++-\rangle +|++-+\rangle+|+-++\rangle
-|-+++\rangle - |---+\rangle +|--+-\rangle +|-+--\rangle - |+---\rangle)/\sqrt{8}\nonumber \\
|T_{1y}^{(2)}\rangle & = & (|+++-\rangle -|++-+\rangle+|+-++\rangle
-|-+++\rangle + |---+\rangle -|--+-\rangle +|-+--\rangle - |+---\rangle)/\sqrt{8}\nonumber \\
|T_{1z}^{(2)}\rangle & = & (|+++-\rangle +|++-+\rangle-|+-++\rangle
-|-+++\rangle + |---+\rangle +|--+-\rangle -|-+--\rangle - |+---\rangle)/\sqrt{8}\nonumber \\
|T_{2x}\rangle & = & (|+++-\rangle -|++-+\rangle-|+-++\rangle
+|-+++\rangle - |---+\rangle +|--+-\rangle +|-+--\rangle - |+---\rangle)/\sqrt{8}\nonumber \\
|T_{2y}\rangle & = & (|+++-\rangle -|++-+\rangle+|+-++\rangle
-|-+++\rangle - |---+\rangle +|--+-\rangle -|-+--\rangle + |+---\rangle)/\sqrt{8}\nonumber \\
|T_{2z}\rangle & = & (|+++-\rangle +|++-+\rangle-|+-++\rangle
-|-+++\rangle - |---+\rangle -|--+-\rangle +|-+--\rangle + |+---\rangle)/\sqrt{8} \nonumber
\end{eqnarray}
\end{widetext}
These kets can be used to block diagonalise the effective 
anisotropic exchange interaction 
on a single tetrahedron.  The $16\times 16$ matrix representing the Hamiltonian
(\ref{HGamma5,1}) is reduced to a 
$3\times 3$ block for the $A_1$ sector:
$\left(\begin{array}{ccc} 6 J_2 & 2\sqrt{3}(-J_1+J_3) & 3 J_4 \\
2\sqrt{3} (-J_1 + J_3)& 4 J_1 - 2J_2 + 4 J_3 &  \sqrt{3} J_4 \\
3 J_4 &  \sqrt{3} J_4 & 6 J_1  \end{array} \right) $, a $2\times 2$ block
for the $A_2$ sector, $\left(\begin{array}{cc}
6 J_2 & 3 J_4 \\
3 J_4 & 6 J_3 \end{array}\right)$, two degenerate $1\times 1$ blocks
for the $E$ sector, 
$-2(J_1+J_2+J_3)$,
three degenerate $2\times 2$ blocks for the $T_2$ sector,
$\left( \begin{array}{cc}
-2 J_2 & - J_4 \\
- J_4 & -2 J_3 \end{array} \right)$, and three degenerate $1\times 1$ blocks
for the $T_2$ sector, $-2J_1$.
These blocks are equivalent to the blocks representing the three-parameter
Hamiltonian 
(\ref{HGamma5,2})
$\left(\begin{array}{ccc} 6 \tilde{J}_2 & 2\sqrt{3}(-J_1+\tilde{J}_3) & 0 \\
2\sqrt{3}(-J_1+\tilde{J}_3) & 4 J_1  - 2 \tilde J_2 +4 \tilde{J}_3 & 0 \\
0 & 0 & 6 J_1  \end{array} \right) $, 
$\left(\begin{array}{cc}
6 \tilde{J}_2 & 0 \\
0 & 6 \tilde{J}_3 \end{array}\right)$,
$-2(J_1+\tilde {J}_2+\tilde{J}_3)$,
$\left( \begin{array}{cc}
-2 \tilde{J}_2 & 0 \\
0 & -2 \tilde{J}_3 \end{array} \right)$, and 
$-2J_1$.

We note that $\Gamma_3$ and $\Gamma_4$ states on single tetrahedron 
yield a different decomposition than for $\Gamma_{5,6}$ states.  In both of these cases, the
decomposition is $A_1\oplus 3 E \oplus 2T_1 \oplus T_2$.  
This leads to a completely different set of degeneracies in the energy
spectrum, and to different linear combinations of
single tetrahedron kets in the eigenstates.


\section{Exchange interaction for $\Gamma_3$ CEF ground state}
We follow the same approach as in Section III to find a general 
effective exchange interaction
for non-Kramers $\Gamma_3$ doublets, such as the CEF ground state of 
Pr in Pr$_2$Sn$_2$O$_7$ (neglecting $J$ mixing):\cite{princep2013} 
\begin{equation}
|\Gamma^{\pm}_{3\rm(Pr)} \rangle = 0.93|\pm 4\rangle \pm0.37| 1\rangle +0.05 |\mp 2\rangle. 
\end{equation}
The operator $\vec{\gamma}$ is defined analogously to ${\vec S}$ and ${\vec \tau}$,
\begin{eqnarray}
\gamma_x & = & |\Gamma_3^{+}\rangle \langle \Gamma_3^{-}| + |\Gamma_3^{-}\rangle \langle \Gamma_3^{+}| \\
\gamma_y & = & -i|\Gamma_3^{+}\rangle \langle \Gamma_3^{-}| +i |\Gamma_3^{-}\rangle \langle \Gamma_3^{+}| \\
\gamma_z & = & |\Gamma_3^{+}\rangle \langle \Gamma_3^{+}| - |\Gamma_3^{-}\rangle \langle \Gamma_3^{-}|. 
\end{eqnarray}
The transformation of $\vec{\gamma}$ under the various symmetry operations is derived
from the transformation of the kets $|\Gamma_3^{\pm}\rangle$.
We find that 
$\gamma_z$ transforms like $\Gamma_2$ of $D_3'$,
while $\gamma_{x,y}$ transforms like 
$\Gamma_3$, the same as for the operators $\vec S$ and $\vec J$.  
However, $\gamma_z$ changes sign under time reversal,
while $\gamma_{x,y}$ does not.
Therefore the effective exchange interaction contains only three terms,\cite{onoda2011}
\begin{eqnarray}
X_{\Gamma_{3},1}  & = &
\sum_{\langle ij\rangle}  \gamma_{iz}\gamma_{jz} \\
X_{\Gamma_{3},2}  
 & = & \frac{1}{3} \sum_{\langle ij \rangle } (\Lambda_{ij}^{*}
\gamma_{i+}  \gamma_{j+} + {\rm h.c.} )   \\
X_{\Gamma_3,3} & = & -\frac{1}{6} \sum_{\langle ij \rangle }(\gamma_{i+} \gamma_{j-}  + {\rm h.c.}).
\end{eqnarray}
These terms are the same as $X_1$, $X_3$ and $X_4$ in $H_{\rm ex}$ (\ref{x1}-\ref{x4});
here $X_2$ is missing because it is not invariant under time reversal.

As noted in the Introduction,
in the restriction to the CEF ground state $|\Gamma_3^{\pm}\rangle$, $J_{x,y}$ vanish while 
$J_z$ is represented by $\sigma_z$.
Non-vanishing operators with matrix elements proportional to $\sigma_x$ are the quadratic
operators $J_z J_x$ and $J_x^2- J_y^2$, while $J_yJ_z$ and $J_x J_y$ yield
$\sigma_y$.  The terms $X_{\Gamma_{3},2}$ and $X_{\Gamma_3,3}$ 
can therefore arise from interactions between nearest neighbour quadrupole moments.  This model has been used to describe Pr spin ice 
pyrochlores.\cite{petit2016,onoda2010,onoda2011,lee2012} 

\section{Summary} 
In modelling exchange interactions in spin systems with symmetry constraints 
there are two aspects that must be considered:  the local site symmetry
(the local CEF) and the global space group symmetry. 
This paper concerns
models for those systems which have doubly degenerate CEF 
ground states, which include many of the rare earth pyrochlores.
Different kinds of CEF doublets may have different symmetries, but each  
may be cast
either as a true spin-1/2 spinor or as a pseudo-spin of some kind. 
The states in a doublet are used
to construct operators that can be represented as the Pauli matrices, whose
symmetry properties are derived from the doublet upon which they are constructed.
These operators act in pairs as nearest neighbour exchange interactions
that are invariant under the global space group symmetry.

The method presented in this paper is an alternative proof of previously proposed
models for the rare earth pyrochlores, and can be generalised to any
magnetic system in which the first excited CEF energy level is well-separated
from the ground state.



\section*{Appendix}

\subsection{Symmetry Considerations}
\begin{figure}[ht]
\includegraphics[height=2.6in]{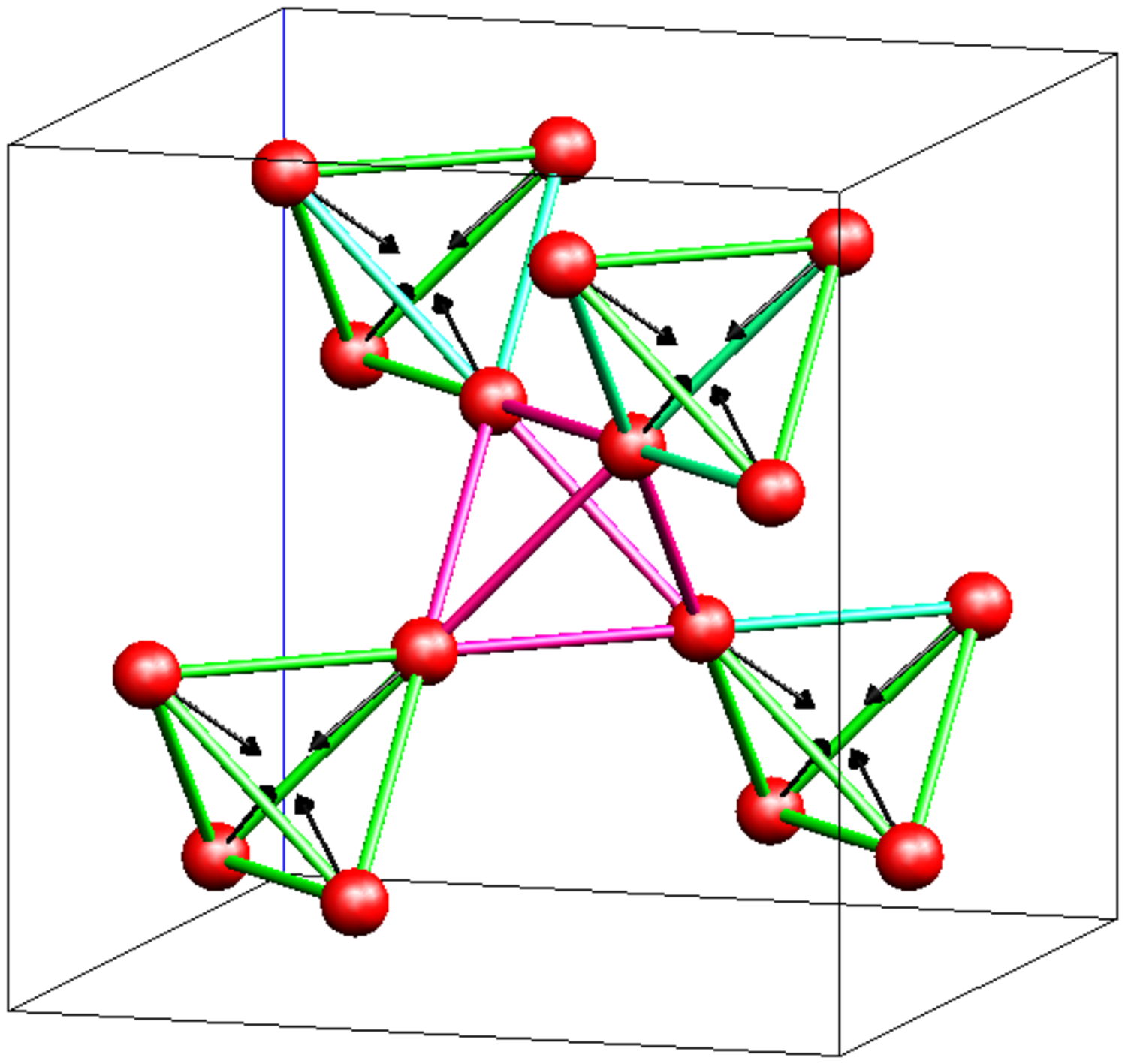}
\includegraphics[height=1.5in]{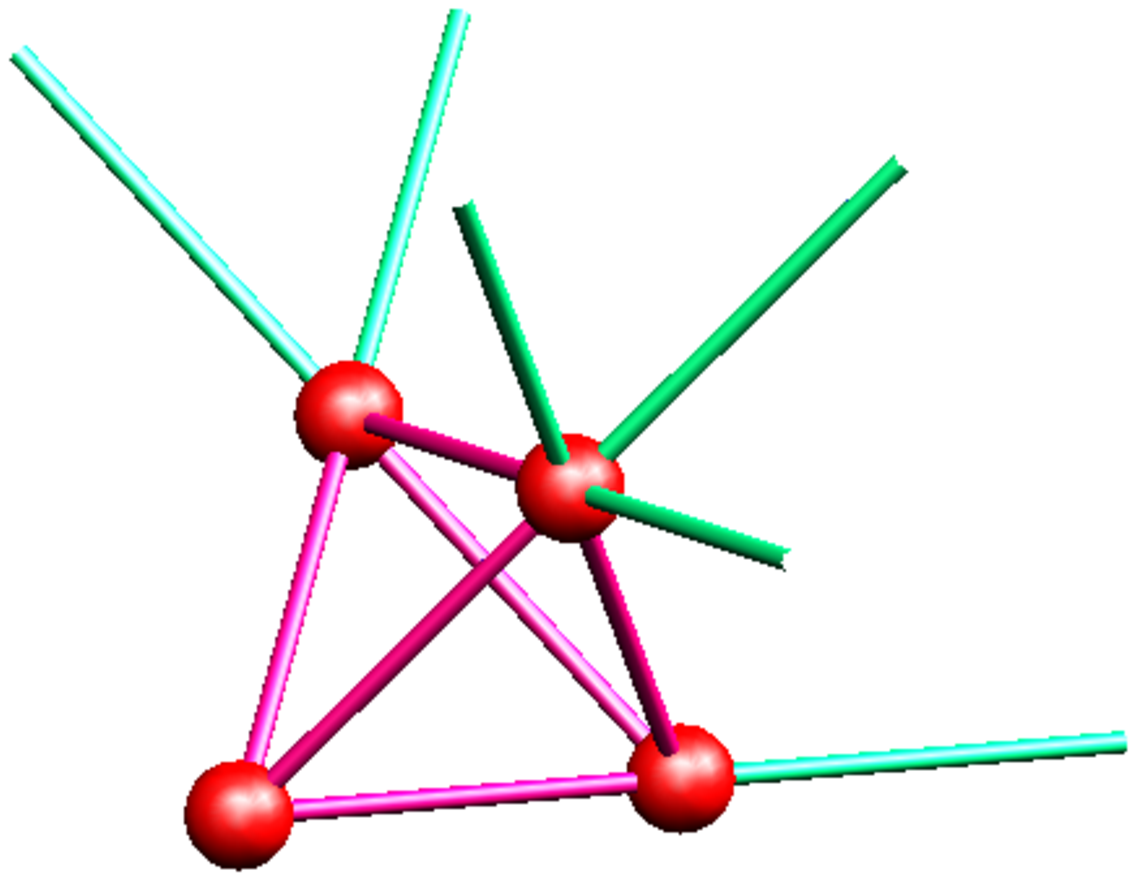}
\caption{(Colour online) The top figure shows  
the rare earth ions at the 16d Wyckoff position and the exchange paths within a cubic cell.  The green/red paths correspond
to type A/B tetrahedra. The black arrows indicate the direction of the local
$z$-axes. 
The bottom figure is a part of the top figure and shows the four rare earth ions and the twelve exchange paths of 
a primitive cell.
}
\label{fig1}
\end{figure}

The magnetic rare earth ions are located at the 16d Wyckoff position of the
cubic
space group $Fd\bar{3}m$.  
Considering Origin Choice I in the International Tables for Crystallography,\cite{space} 
the four sites of a primitive unit cell 
are located at the positions listed in Table \ref{wyckoff}.
The remaining twelve sites found in the cubic cell are obtained by fcc lattice
translations. The cubic cell and the primitive cell are shown in Fig.\ \ref{fig1}.
The underlying point group symmetry is the octahedral group $O_h$, 
which includes 2-fold rotations and 4-fold screw rotations about the cubic axes, 
3-fold rotations about
the cube diagonals (shown as black arrows in the figure), and 2-fold screw rotations about the $[110]$ directions, as well as inversion centres at each rare earth
site.

\begin{table}
\begin{tabular}{c|c|c}
\hline
Site \# & Position & 3-fold axis \\
\hline
1 & $(5/8,5/8,5/8)$ & $(1,1,1)$ \\
2& $(3/8,3/8,5/8)$ & $(-1,-1,1)$ \\
3 & $(3/8,5/8,3/8)$ & $(-1,1,-1)$ \\
4 & $(5/8,3/8,3/8)$ & $(1,-1,-1)$  \\
\hline
\end{tabular}
\caption{Locations of the four sites of the 16d Wyckoff position. \label{wyckoff}}
\end{table}

It is convenient to adopt a local coordinate system for each of the four sites
of the 16d Wyckoff position. Global coordinates are denoted by superscripts 
while the local coordinates are denoted by subscripts.  
The angular momentum operators for local coordinates are given in terms of
the global coordinates as follows:
$$ \begin{array}{ll}
J_{1x} = (J_1^x + J_1^y - 2 J_1^z)/\sqrt 6, & J_{2x} = (-J_2^x -J_2^y -2J_2^z)/\sqrt{6}, \\ 
J_{1y} = (-J_1^x + J_1^y)/\sqrt{2}, & J_{2y} = (J_2^x - J_2^y)/\sqrt{2},\\
J_{1z} = (J_1^x + J_1^y + J_1^z)/\sqrt{3}, &  J_{2z} = (-J_2^x - J_2^y + J_2^z)/\sqrt 3, \\
J_{3x} = (-J_3^x + J_3^y + 2 J_3^z)/\sqrt 6, & J_{4x} = (J_4^x -J_4^y +2J_4^z)/\sqrt{6}, \\     
J_{3y} = (J_3^x + J_3^y)/\sqrt{2}, & J_{4y} = (-J_4^x - J_4^y)/\sqrt{2}, \\
J_{3z} = (-J_3^x + J_3^y - J_3^z)/\sqrt{3}, &  J_{4z} = (J_4^x - J_4^y - J_4^z)/\sqrt 3. \end{array}$$
The local $z$-axes are the 3-fold symmetry axes of the crystal, and in particular
they are 3-fold symmetry axes for the 16d Wyckoff position.
The local $z$-axes are shown as black arrows in Fig.\ \ref{fig1}.
Different choices of the local $x$ and $y$ axes are possible, provided
they are perpendicular to $z$ and obey the right hand rule.

The 16d positions form a corning-sharing tetrahedral lattice.  The tetrahedra
alternate between two orientations, which we label A and B.  Each site is
on a corner that is shared between an A tetrahedron and a B tetrahedron.  (We can
assume 
that the  sites listed in Table \ref{wyckoff} 
are the corners of a B tetrahedron.) 

Alternatively,  the corner-sharing tetrahedral lattice 
can be defined by the edges of the tetrahedra. Each edge belongs to either an
A or a B tetrahedron (the edges are not shared).
Nearest-neighbour interactions between rare earth sites are positioned 
along paths which connect the sites, i.e., the edges of tetrahedra, 
therefore, they 
are uniquely associated with either an A type or B type tetrahedron.  

The site symmetry of the 16d position is $D_{3d}$, where the 
three-fold axes point in the $[111]$ directions of the crystal.
The inversion element of $D_{3d}$ is usually handled separately, because within a given
$J$ manifold the orbital angular momentum $L$ is fixed, and is simply either even or odd
under inversion.  
However, the double set of rotations must be considered for half-integral $J$, so
the relevant point group for the 16d positions is the double group $D_3'$.

\begin{table}
\begin{tabular}{l|rrrrrr}
\hline
$D_3'$ & $E$ & $R$ & $2C_3$ & $2RC_3$ & $3C'_2$ & $3RC_2$ \\
\hline
$\Gamma_1$ & 1 & 1 & 1 & 1 & 1 & 1 \\
$\Gamma_2$ & 1 & 1 & 1 & 1 & -1 & $-1$ \\
$\Gamma_3$ & 2 & 2 & $-1$ & $-1$ & 0 & 0 \\
\hline
$\Gamma_4$ & 2 & $-2$ & 1 & $-1$ & 0 & 0 \\
$\Gamma_{5,6}$ & 1 & $-1$ & $-1$ & $1$ &$i$  & $-i$ \\
           & 1 & $-1$ & $-1$ & $1$ & $-i$ & $i$ \\
\hline
\end{tabular}
\caption{The character table of the double group $D'_3$.
\label{D3}}
\end{table}

Table \ref{D3} shows the character table for the double group $D'_3$.  The CEF levels of ions at the 16d positions can be classified by these representations.
For integral $J$ ions, there are
three representations, two singlets and a doublet.
For half-integral $J$, (which change sign under rotations by 
$2\pi$), there are 
two representations, that are necessarily (by Kramers Theorem) two
dimensional.

\begin{table}
\begin{tabular}{c|ccccccc}
\hline
  & $C_{2z}$ & $C_{2y}$ & $C_{3[111]}$ & $C_{2[110]}$ & $C_{4z}$ & $I$ & ${\cal K}$ \\
\hline
1 & 2 & 3 & 1 & 2 & 4 & 1 & 1 \\
2 & 1 & 4 & 4 & 1 & 3  & 2& 2 \\
3 & 4 & 1 & 2 & 3 & 1 & 3 & 3 \\
4 & 3 & 2 & 3 & 4 & 2 & 4 & 4 \\
\hline
A/B & A/B & A/B & A/B & B/A & B/A & B/A & A/B \\
\hline
$J^x$ & $-J^x$ &  $-J^x$ & $J^{y} $ &  $J^y$ &  $J^y$ & $J^x$& $-J^x$  \\
$J^y$ & $-J^y$ & $J^y$ & $J^{z}$ &  $ J^x$ & $-J^x$ & $J^y$ & $-J^y$  \\
$J^z$ & $J^z$ & $-J^z$ & $J^x$ & $-J^z$ & $J^z$ & $J^z$ & $-J^z$ \\
\hline
$J_x$ & $J_x$ &  $J_x$ & $J_{x'} $ &  $-J_x$ &  $-J_x$ & $J_x$ & $-J_x$  \\
$J_y$ & $J_y$ & $J_y$ & $J_{y'}$ &  $ J_y$ & $J_y$ & $J_y$ & $-J_y$ \\
$J_z$ & $J_z$ & $J_z$ & $J_z$ & $-J_z$ & $-J_z$ & $J_z$ & $-J_z$ \\
\hline
$\beta_x$ & $\beta_x$ &  $\beta_x$ & $\beta_{x} $ &  $-\beta_x$ &  $-\beta_x$ & $\beta_x$ 
& $-\beta_x$ \\
$\beta_y$ & $\beta_y$ & $\beta_y$ & $\beta_{y}$ &  $ -\beta_y$ & $-\beta_y$ & $\beta_y$ & $-\beta_y$  \\
$\beta_z$ & $\beta_z$ & $\beta_z$ & $\beta_z$ & $\beta_z$ & $\beta_z$ & $\beta_z$ & $-\beta_z$  \\
\hline
$\gamma_x$ & $\gamma_x$ &  $\gamma_x$ & $\gamma_{x'} $ &  $-\gamma_x$ &  $-\gamma_x$ & $\gamma_x$
& $\gamma_x$ \\
$\gamma_y$ & $\gamma_y$ & $\gamma_y$ & $\gamma_{y'}$ &  $ \gamma_y$ & $\gamma_y$ & $\gamma_y$ & $\gamma_y$  \\
$\gamma_z$ & $\gamma_z$ & $\gamma_z$ & $\gamma_z$ & $-\gamma_z$ & $-\gamma_z$ & $\gamma_z$ & $-\gamma_z$  \\
\hline
\end{tabular}
\caption{Transformation of site numbers,
tetrahedron type, and various operators under the space group generators and time reversal. 
$x' = -x/2 + \sqrt 3 y/2$, $y' = -y/2 - \sqrt{3} x/2$.
\label{spintable}}
\end{table}

Table \ref{spintable} shows how different sites, tetrahedra, and operators transform under the 
space group operations and time reversal ${\cal K}$.  
The operations are 2-fold, 3-fold and 4-fold rotations 
around a global axis specified by the subscript and inversion $I$. Some of
the rotations are screw rotations (for details see Ref.\ \onlinecite{space}).
For example, under a $C_{4z}$ screw rotation, $J_{1x} \rightarrow -J_{4x}$
and type A paths become type B paths.

\begin{table*}
\begin{tabular}{|l|rrrrr|rrrrr||cccccccccc|}\hline
 $O_h$ & $E$ & $C_3$ &  $C_2$ & $C_2^{'}$ & $C_4$&  $I$ & $IC_3$ &  $IC_2$ & $IC_2^{'}$ & $IC_4$ 
& $A_{1g}$ & $A_{2g}$ &$ E_g$ & $T_{1g}$ &  $T_{2g}$ & $A_{1u}$ &$A_{2u}$ & $E_u$
& $ T_{1u}$  & $T_{2u}$ \\ \hline
$J_{iz}J_{jz}$ & 12 &0 &4&0& 0& 0 &0 &0&4&0& 1 &- & 1 &-&  1 &- & 1 & 1 & 1 & -    \\
$\begin{array}{l} J_{iz}J_{jx}, \\ J_{iz}J_{jy} \end{array} $ &  48 &0 &0&0&0 &0&0&0&0&0&   1 &1 & 2 & 3 & 3 & 1 & 1 & 2 & 3 & 3 \\
$\begin{array}{l}
J_{ix}J_{jx}, \\ J_{iy}J_{jy} \end{array}$ & 24 &0 &8&0& 0 & 0&0&0&8 &0&
2 & -  & 2 &- & 2 &-  & 2 & 2 & 2 &- \\
$J_{ix}J_{jy}$ & 24 & 0 & 0 & 0 & 0 &0&0&0&-4&0&
- & 1 & 1 & 2 & 1 & 1 &-  & 1 & 1 & 2 \\
\hline
total & 108 & 0 & 12 & 0 & 0 &0&0&0&8&0& 
4 & 2 & 6 & 5 & 7 & 2 & 4 & 6 & 7 & 5 \\
\hline
$J_{i\pm}J_{j\pm} $ & 24 &0 &8&0& 0 & 0&0&0&0 &0&
1 & 1 & 2 & 1 & 1 & 1 & 1 & 2 & 1 & 1 \\
$J_{i+}J_{j-}$ & 24 & 0 & 0 & 0 & 0 &0&0&0&4&0& 
1 &- & 1 & 1 & 2 &- & 1 & 1 & 2 & 1 \\
\hline
\hline
$\beta_{iz}\beta_{jz}$ & 12 &0 &$4$ &0& 0& 0 &0 &0&$4 $&0&
1 &-& 1 &-&1 &-&1 & 1 & 1 &- \\
$ \beta_{iz}\beta_{jx,y}$ &  48 &0 &0&0&0 &0&0&0&-8&0&
- & 2 & 2 & 4 & 2 & 2 &- & 2 & 2 & 4 \\
$ \beta_{ix}\beta_{jx}$ & 12 &0 &$4 $&0& 0 & 0&0&0&4 &0&
1 &-& 1 &-&1 &-&1 & 1 & 1 &- \\
$ \beta_{iy}\beta_{jy}$ & 12 &0 &$4 $&0& 0 & 0&0&0&4 &0& 
1 &-& 1 &-&1 &-&1 & 1 & 1 & - \\
$\beta_{ix}\beta_{jy}$ & 24 & 0 & 0 & 0 & 0 &0&0&0&4 &0& 
1 & -& 1 & 1 & 2 &- & 1 & 1 & 2 & 1 \\
\hline
total & 108 & 0 & 12 & 0 & 0 &0&0&0&8&0& 
4 & 2 & 6 & 5 & 7 & 2 & 4 & 6 & 7 & 5 \\
\hline
$\beta_{i\pm}\beta_{j\pm} $ & 24 &0 &8&0& 0 & 0&0&0&8 &0&
2 &-& 2 &- & 2 &- & 2 & 2 & 2 &- \\
$\beta_{i+}\beta_{j-}$ & 24 & 0 & 0 & 0 & 0 &0&0&0&4&0& 
1 &- & 1 & 1 & 2 &- & 1 & 1 & 2 & 1 \\
\hline
\end{tabular}
\caption{
Characters associated
with each type of bilinear spin operator.  The first part of the
top row lists the the classes of the point group $O_h$; the
second part lists the irreducible representations of $O_h$.
The first column lists the different types of bilinears.  The left
hand-side array of numbers
is the characters, and the right hand-side array of numbers
gives the
representation decomposition for
each type of bilinear.  
The $C_2$ operations are rotations about the main cubic axes (the $[100]$ directions),
and $C_2'$ are rotations about the $[110]$ directions.
  \label{charac-spin} }
\end{table*}

Using the results in Table \ref{spintable}, the characters associated with
each kind of bilinear spin operator can be calculated, as listed in 
Table \ref{charac-spin}.  Each bilinear operator is associated with two 
nearest neighbour sites within a unit cell. There are twelve such
pairs altogether: six pairings of the four sites, and two of each of these
corresponding to A-type and B-type paths. 
The first column of characters (under the identity
operation $E$)
simply counts the number of operators of a given type. 
Among the remaining operations, only those that  preserve
the path (A vs B) of the bilinear
can yield a non-zero character.  These are
$C_3$, $C_2$, $IC_2'$ and $IC_4$.  Among these, $C_3$ and $IC_4$
have vanishing characters because $C_3$ permutes three sites and $IC_4$
permutes four sites (so the site numbers on the bilinear must change).  
Therefore the only classes with 
non-vanishing characters are $C_2$ and $IC_2'$. 
These characters can be calculated by considering the 
operations $C_{2z}$, $C_{2[110]}$ and $I$ in Table \ref{spintable}
for $\vec J$ and $\vec \beta$ ($\vec \gamma$ transforms the same way as 
$\vec J$ under the space group operations, but differs under time reversal,
as discussed in Section IV).

The entries under the column 
labelled $A_{1g}$ in Table \ref{charac-spin} are the 
number of 
bilinear invariants for the pyrochlore crystal.
Each invariant corresponds to a term in the Hamiltonian.

\subsection{The anisotropic exchange interaction}
The exchange interaction written as (\ref{x1}-\ref{x4}) is a useful form for
analytic calculations.  
Alternatively, 
the four terms in the Hamiltonian can be written as
\begin{widetext}
\begin{eqnarray}
-3 X_1 & =  & \sum_{\rm tetra} J_{1z}J_{2z} + J_{1z}J_{3z} + J_{1z}J_{4z} + J_{2z}J_{3z} + J_{2z}J_{4z} + J_{3z}J_{4z} +  \\
-\frac{3}{\sqrt{2}}X_2 & = & \sum_{\rm tetra} J_{1z}J_{2x} + J_{1x}J_{2z} +J_{3z}J_{4x} +  J_{3x}J_{4z} 
+ J_{1z}J'_{4x} + J'_{1x}J_{4z} \nonumber  \\
& & + J_{1z}J''_{3x}  
+ J''_{1x}J_{3z}
+J_{2z}J'_{3x} + J'_{2x}J_{3z} + J_{2z}J''_{4x} + J''_{2x}J_{4z} \\
\frac{3}{4}X_1 - \frac{3}{2}X_2   & = & \sum_{\rm tetra} J_{1x}J_{2x} + J_{3x}J_{4x} + J'_{1x}J'_{4x}
+J''_{1x}J''_{3x} + J'_{2x}J'_{3x} +J''_{4x}J''_{2x}  \\
-\frac{3}{4}X_3 - \frac{3}{2}X_2 & = & \sum_{\rm tetra} J_{1y}J_{2y} + J_{3y}J_{4y} + J'_{1y}J'_{4y}
+J''_{1y}J''_{3y} + J'_{2y}J'_{3y} +J''_{4x}J''_{2y} 
\end{eqnarray}
\end{widetext}
where $x' = -\frac{x}{2} + \frac{\sqrt 3y}{2}$, 
$y' = -\frac{y}{2} - \frac{\sqrt  3x}{2}$,
$x'' = -\frac{x}{2}- \frac{\sqrt 3y}{2}$, and
$y'' = -\frac{y}{2}+ \frac{\sqrt 3x}{2}$.
The sums are over all tetrahedra (both A and B orientations) in the 
pyrochlore lattice.  This allows us to compare the terms in 
this Hamiltonian with the terms of the $\Gamma_{5,6}$ model (\ref{betaz}-\ref{betaxy}): 
in this Hamiltonian some of the operators are rotated by 120 degrees and by 240 degrees in 
the $xy$-plane, 
while in the $\Gamma_{5,6}$ model, all of the operators are
rotated by the same angle~$\theta$.

The exchange interaction can also be written using global axes:
\begin{widetext}
\begin{eqnarray}
X_{s1} & = & \sum_{\rm tetra} J_{1}^zJ_{2}^z + J_{1}^yJ_{3}^y+J_{1}^xJ_{4}^x+J_{2}^xJ_{3}^x+J_{2}^yJ_{4}^y +J_{3}^zJ_{4}^z  \\
 & = & - X_1 + \frac{1}{2}X_2 + \frac{1}{2} X_3 - X_4 \\
X_{s2} & = & \sum_{\rm tetra} J_{1}^xJ_{2}^x +J_1^yJ_2^y +  J_{1}^xJ_{3}^x+J_1^zJ_3^z + J_{1}^yJ_{4}^y+J_1^zJ_4^z +J_{2}^yJ_{3}^y + J_2^zJ_3^z+J_{2}^xJ_{4}^x+J_2^zJ_4^z  +J_{3}^xJ_{4}^x+J_1^yJ_4^y  \\
& = & 2X_1 + \frac{1}{2}X_2 + \frac{1}{2} X_3 + 2 X_4 \\
X_{s3}  & = & \sum_{\rm tetra} J_{1}^xJ_{2}^y + J_1^yJ_2^x +J_1^x J_3^z + J_1^z J_3^x +  J_1^yJ_4^z + J_1^zJ_4^y - J_2^yJ_3^z - J_2^z J_3^y - J_2^x J_4^z - J_2^z J_4^x - J_{3}^xJ_{4}^y - J_3^yJ_4^x   \\
 & = & 2 X_1 + \frac{1}{2} X_2 - X_3 - X_4 \\
X_a & = & \sum_{\rm tetra} [\vec{J}_1\times\vec{J}_2]^x - [\vec{J}_1\times\vec{J}_2]^y +[\vec{J}_1\times\vec{J}_3]^z - [\vec{J}_1\times \vec{J}_3]^x +[\vec{J}_1\times\vec{J}_4]^y -[\vec{J}_1\times \vec{J}_4]^z \nonumber \\ 
& & -[\vec{J}_2\times\vec{J}_3]^z - [\vec{J}_2\times \vec{J}_3]^y +[\vec{J}_2\times\vec{J}_4]^x +[\vec{J}_2\times \vec{J}_4]^z -[\vec{J}_3\times\vec{J}_4]^x -[\vec{J}_3\times \vec{J}_4]^y  \\
& = & -4 X_1 + \frac{1}{2}X_2 -X_3 + 2 X_4
\end{eqnarray}
\end{widetext}
Here, the sum of the first two terms $X_{s1}+ X_{s2}$ yields the isotropic exchange interaction.  The third term is another completely symmetric contribution while the fourth is completely anti-symmetric.

\vspace{.1in}
This work was supported by NSERC.

\end{document}